\documentclass[12pt,a4paper]{article}

\usepackage{epsfig}
\usepackage{amsmath,amsfonts,amssymb}
\usepackage{t1enc}
\usepackage{cite}

\newcommand{\ptlep}{p_t^\mathrm{lep}}
\newcommand{\ptmax}{p_t^{j,\mathrm{max}}}
\newcommand{\ptbmax}{p_t^{b,\mathrm{max}}}
\newcommand{\ptmiss}{p_t\!\!\!\!\!\!\!\! \not \,\,\,\,\,\,}
\newcommand{\mthad}{m_T^\mathrm{had}}
\newcommand{\mtlep}{m_T^\mathrm{lep}}
\providecommand{\openone}{\leavevmode\hbox{\small1\kern-3.8pt\normalsize1}}

\begin{document}

\begin{center}
\begin{Large}
{\bf Pair production of heavy $\boldsymbol{Q=2/3}$ singlets at LHC}
\end{Large}

\vspace{0.5cm}
J. A. Aguilar--Saavedra \\[0.2cm] 
{\it Departamento de Física and CFTP, \\
Instituto Superior Técnico, P-1049-001 Lisboa, Portugal} \\
\end{center}

\begin{abstract}
We examine the LHC discovery potential for new $Q=2/3$ quark singlets $T$ in the
process $gg,qq \to T \bar T \to W^+ b \, W^- \bar b$, with one $W$ boson
decaying hadronically and the other one leptonically. A particle-level
simulation of this signal and its main backgrounds is performed, showing that 
heavy quarks with masses of 500 GeV or lighter can be discovered at the $5
\sigma$ level after a few months of running, when an integrated luminosity of 3
fb$^{-1}$ is collected. With a luminosity of  100 fb$^{-1}$, this process can
signal the presence of heavy quarks with masses up to approximately 1 TeV.
Finally, we discuss the complementarity among $T \bar T$, $Tj$ production and
indirect constraints from precise electroweak data in order to discover a new
quark or set bounds on its mass.
\end{abstract}

\section{Introduction}

The Large Hadron Collider (LHC) will be a powerful tool to explore energies up
to the scale of a few TeV. It is expected to provide some
striking evidence of new physics, for instance of a light Higgs boson, in its
first months of operation \cite{atlascms,lhcilc}.
Among many promising possibilities for the discovery
of new particles, LHC will offer an ideal environment for the
production of heavy quarks.
New quarks of either charge can be copiously
produced in pairs through QCD
interactions, namely via gluon fusion and quark-antiquark annihilation, if there
is available phase space \cite{eichten,frampton}. Up-type quarks $T$ can also
be produced in association with light jets, {\em e.g.} in the
processes $q b \to q' T$, $\bar q' b \to \bar q T$
(here and throughout this Letter $q=u,c$, $q'=d,s$),
provided their mixing with the bottom quark is sizeable. New interactions may
also bring about further production mechanisms. 
The prospects for heavy quark detection depend on the production processes (with
their respective cross sections) as well as on the decay modes (and their
relevant backgrounds), which are
distinctive of the Standard Model (SM) extension considered.

The presence of a fourth sequential generation is disfavoured by
naturalness arguments\footnote{For a fourth quark generation, anomaly
cancellation requires the simultaneous presence of a lepton doublet. LEP
measurement of the $Z$ invisible width sets the number of light neutrino
species to three, and additional neutrinos must be heavier than 45 GeV
\cite{pdb}, in sharp contrast with the smallness of the light neutrino masses
$m_\nu \lesssim 1$ eV.}
and precision electroweak data, which leave a small window for the new quark
masses consistent with the experimental measurement of the $\mathrm{S}$,
$\mathrm{T}$, $\mathrm{U}$
parameters \cite{pdb}. On the other hand, heavy $\mathrm{SU}(2)_L$ quark
singlets with charges $Q=2/3$ or $Q=-1/3$ can exist with a moderate mixing of
order $10^{-2}-10^{-1}$
with the SM quarks. Here we are concerned with the first possibility.
Models with large extra dimensions with for
example $t_R$ in the bulk predict the existence of a tower of $Q=2/3$ singlets
$T_{L,R}^{(n)}$. If there is multilocalisation the lightest one, $T_{L,R}^{(1)}$
can have a mass of 300 GeV or larger, and a sizeable mixing with the top quark
\cite{jose}. (The class of extra-dimensional models having a light $Q=2/3$
singlet mixing with the top quark is enlarged when corrections localised on the
branes to the kinetic terms of fermions and bosons are taken into account
\cite{bkt}.) Little Higgs models \cite{lhiggs} include in their additional
spectrum an up-type singlet, which is expected to have a mass of 1 TeV or
larger. Quark singlets  also appear in some grand unified theories
\cite{frampton,barger}. Their effects in low energy and top physics have 
already been studied \cite{largo}. In this Letter we address their direct
observation at LHC through pair production $gg,qq \to T \bar T$ \cite{paco2}.

We note that for heavy quark masses $m_T \gtrsim 800$ GeV and a coupling to the
bottom quark $V_{Tb}$ of the size suggested by the experimental measurement of
the $\mathrm{T}$ parameter, single $T$ production $pp \to Tj$ has a
larger cross section than pair production and can then explore larger mass
scales \cite{azuelos,costanzo}. Therefore,
$Tj$ production will eventually set more stringent limits (albeit
dependent on $V_{Tb}$) on heavy quark
masses if a positive signal is not observed. However, two important points have
to be remarked: ({\em i\/}) the $Tj$ cross section is proportional
to $|V_{Tb}|^2$, hence for small mixings this process becomes less relevant;
({\em ii\/}) pair production has the best sensitivity to the
presence of new quarks having masses of several hundreds of GeV. If new quarks
exist in this mass range, $T \bar T$ production would allow to observe a signal
in a rather short time.

In the following we briefly review the mixing of the new quark, its interactions
and decay modes. After summarising the relevant aspects of the signal and
background generation, we will
present our results for quark masses of 500 GeV and 1 TeV. Finally,
the relation between $T \bar T$, $Tj$ production and indirect constraints from
the $\mathrm{T}$ parameter will be discussed.

\section{SM extensions with $\boldsymbol{Q=2/3}$ singlets}
\label{sec:2}

The addition of two $\mathrm{SU}(2)_L$ singlet fields $T^0_{L,R}$ to the quark
spectrum modifies the weak and scalar interactions involving $Q=2/3$ quarks.
(We denote weak eigenstates with a zero superscript, to distinguish them from 
mass eigenstates which do not bear superscripts.)
Using standard notation, these interactions read
\begin{eqnarray}
\mathcal{L}_W & = & - \frac{g}{\sqrt 2} \left[ \bar u \gamma^\mu V P_L  d
  \; W_\mu^+ + \bar d \gamma^\mu V^\dagger P_L u \; W_\mu^- \right] \,,
  \nonumber \\
\mathcal{L}_Z & = & - \frac{g}{2 c_W} \bar u \gamma^\mu \left[ X P_L
  - \frac{4}{3} s_W^2 \openone_{4 \times 4}  \right] u \; Z_\mu \,, \nonumber \\
\mathcal{L}_H & = & \frac{g}{2 M_W} \,
\bar u \left[ \mathcal{M}^u X P_L  + X \mathcal{M}^u P_R \right] u
 \;  H \,,
\label{ec:1}
\end{eqnarray}
where $u=(u,c,t,T)$, $d=(d,s,b)$ and $P_{R,L} = (1 \pm \gamma_5)/2$.
The extended Cabibbo-Kobayashi-Maskawa (CKM) matrix $V$ is of dimension
$4 \times 3$, $X = V V^\dagger$ is a non-diagonal $4 \times 4$ matrix and
$\mathcal{M}^u$ is the $4 \times 4$ diagonal up-type quark mass matrix.
The new mass eigenstate $T$ is expected to couple mostly with third
generation quarks $t$, $b$, because $T_L^0$, $T_R^0$ preferrably mix with
$t_L^0$, $t_R^0$, respectively, due to the large top quark mass. $V_{Tb}$ is
mainly constrained by the contribution of the new quark to the $\mathrm{T}$
parameter \cite{silva,largo},
\begin{equation}
\mathrm{T} = \frac{N_c}{16\pi s_W^2 c_W^2} \left\{ |V_{Tb}|^2 \left[ 
\theta_+(y_T,y_b) - \theta_+(y_t,y_b) \right] 
- |X_{tT}|^2 \theta_+(y_T,y_t) \right\} \,,
\label{ec:T}
\end{equation}
where $N_c=3$ is the number of colours, $y_i = (\overline{m_i}/M_Z)^2$, 
$\overline{m_i}$ being the $\overline{\mathrm{MS}}$ mass of the quark $i$ at the
scale $M_Z$, $|X_{tT}|^2 \simeq |V_{Tb}|^2 (1-|V_{Tb}|^2)$ and \cite{silva}
\begin{equation}
\theta_+(y_1,y_2) = y_1 + y_2 -\frac{2 y_1 y_2}{y_1-y_2} \log
\frac{y_1}{y_2} \,.
\end{equation}
The experimental measurement $\mathrm{T} = 0.12 \pm 0.10$ \cite{lepewwg},
obtained setting $\mathrm{U}=0$, implies $\mathrm{T} \leq 0.28$ with a 95\%
confidence level (CL), and the corresponding limit $|V_{Tb}| \leq 0.26-0.18$
for $m_T = 500-1000$ GeV (see also Ref.~\cite{roberto}).\footnote{The
new quark contribution to $\mathrm{U}$ is much smaller, of order $10^{-2}$
\cite{largo}, thus it makes sense using this value for $\mathrm{T}$. If we
take $\mathrm{T} = -0.17 \pm 0.12$ \cite{pdb} the limits obtained
are much stronger, $\mathrm{T} \leq 0.027$ and thus $|V_{Tb}| \leq 0.08-0.06$
for $m_T = 500-1000$ GeV. We will consider both possibilities in our analysis.}
Mixing of $T_L^0$ with $u_L^0$, $c_L^0$, especially with the latter, is very
constrained by parity violation experiments and the measurement of $R_c$ and
$A_\mathrm{FB}^{0,c}$ at LEP, respectively \cite{london,prl},
implying small $X_{uT}$, $X_{cT}$.
The charged current couplings with $d,s$ must be small as well,
$|V_{Td}|, |V_{ts}| \sim 0.05$, because otherwise the new quark would
give large loop contributions to kaon and $B$ physics observables \cite{largo}.
Therefore,
$|V_{Td}|,|V_{Ts}| \ll |V_{Tb}|$ and
$|X_{uT}|, |X_{cT}| \ll |X_{tT}|$. In specific models there may be additional
interactions, {\em e.g.} mediated by new gauge bosons, giving further
contributions to experimental observables. These extra terms
might (partially) cancel the
ones from the new quark, loosening the constraints on its
couplings. Although new interactions may modify the allowed range for $V_{Tb}$,
it is unlikely that they alter the above hierarchy. Then, the relevant decays
of the new
quark are $T \to W^+ b ,\, Zt ,\, Ht$, with partial widths
\begin{eqnarray}
\Gamma(T \to W^+ b) & = & \frac{\alpha}{16 \, s_W^2} |V_{Tb}|^2
  \frac{m_T^3}{M_W^2} \left[ 1-3 \frac{M_W^4}{m_T^4} + 2 \frac{M_W^6}{m_T^6}
  \right] \,, \nonumber \\
\Gamma(T \to Z t) & = & \frac{\alpha}{16 s_W^2 c_W^2} |X_{tT}|^2
  \frac{m_T^2}{M_Z^2} f(m_T,m_t,M_Z) \nonumber \\
  & & \times  \left[ 1 + \frac{M_Z^2}{m_T^2}
  - 2  \frac{m_t^2}{m_T^2} - 2  \frac{M_Z^4}{m_T^4}  + \frac{m_t^4}{m_T^4}
  + \frac{M_Z^2 m_t^2}{m_T^4} \right] \,, \nonumber \\
\Gamma(T \to H t) & = & \frac{\alpha}{16 s_W^2} |X_{tT}|^2
 \frac{m_T^2}{M_W^2} f(m_T,m_t,M_H) \nonumber \\
  & & \times  \left[ 1 + \frac{3}{4} \frac{m_t}{m_T} - \frac{1}{2}
  \frac{m_t^2}{m_T^2} - \frac{M_H^2}{m_T^2} + \frac{3}{4} \frac{m_t^3}{m_T^3}
  + \frac{m_t^4}{m_T^4} - \frac{m_t^2 M_H^2}{m_T^4} \right] \,.
\label{ec:3}
\end{eqnarray}
The kinematical function
\begin{equation}
f(m_T,m_t,M) \equiv \frac{1}{2 m_T} (m_T^4 + m_t^4 + M^4 - 2 m_T^2 m_t^2 
- 2 m_T^2 M^2 - 2 m_t^2 M^2)^{1/2}
\end{equation}
approximately equals $m_T/2$ for $m_T \gg m_t,M$. For a heavy $T$ and a light
Higgs, we have
$\mathrm{Br}(T \to W^+ b) \simeq 0.5$,
$\mathrm{Br}(T \to Z t) \simeq 0.25$, $\mathrm{Br}(T \to H t) \simeq 0.25$.
To our knowledge, there are not experimental searches for new $Q=2/3$ quarks
giving lower bounds for their masses. However, for the expected production cross
sections and decay branching ratios, it seems that quarks with masses around 200
GeV ought to be visible with present Tevatron Run II data \cite{tev2}.

The decays $T \to Zt \to \ell^+ \ell^- W^+ b$,
$\ell=e,\mu$ give 
a cleaner final state than $T \to W^+ b$, but with a branching ratio 30 times
smaller. It has already been found that the channel $T \to W^+ b$
($\bar T \to W^- \bar b$), with $W \to \ell \nu$, gives the best discovery
potential in single $T$ production \cite{azuelos}. In $T \bar T$ production
we select the final states $T \bar T \to W^+ b \, W^- \bar b$, with one $W$
boson decaying leptonically and the other one hadronically.
The larger cross section in this decay
mode allows to obtain a better statistical significance for the signal, while
the backgrounds can be greatly reduced with kinematical cuts.

\section{Signal and background simulation}
\label{sec:3}

The main backgrounds for the $T \bar T$ signal
\begin{equation}
gg,qq \to T \bar T \to W^+ b \, W^- \bar b \to \ell^+ \nu b \, \bar q q' \bar b
\;, \quad \quad \ell=e,\mu
\label{ec:sig}
\end{equation}
are given by $t \bar t$, $W b \bar b jj$, $Z b \bar b jj$ and $t \bar b j$
production, 
\begin{align}
& gg,qq \to t \bar t \to W^+ b \, W^- \bar b \to \ell^+ \nu b \, \bar q q'
\bar b \,, \nonumber \\
& pp \to W b \bar b jj \to \ell \nu b \bar b jj \,, \nonumber \\
& pp \to Z b \bar b jj \to \ell^+ \ell^- b \bar b jj \,, \nonumber \\
& pp \to t \bar b j \to W^+ b \bar b j \to \ell^+ \nu b \bar b j \,.
\label{ec:bkg}
\end{align}
The charge conjugate processes are understood to be summed in all cases.
In the background evaluation we do not consider final states with $\tau$ leptons
(which can decay leptonically $\tau \to e \nu_\tau \bar \nu_e$,
$\tau \to \mu \nu_\tau \bar \nu_\mu$) because the electrons and muons produced
in $\tau$ decays are softer, and in our analysis we eventually
require $e,\mu$ with high transverse momenta. $Wjjjj$ and $Zjjjj$ production are
reduced to negligible levels with the requirement of two $b$ tags, which
suppresses their contributions by a factor $\sim 10^{-4}$. The signal
and the $t \bar t$, $t \bar b j$ backgrounds are evaluated with
our own Monte Carlo generators, including
all finite width and spin effects. We calculate the matrix elements using {\tt
HELAS} \cite{helas} with running coupling constants evaluated at the scale of
the heavy quark $T$ or $t$. $W b \bar b jj$  and $Z b \bar b jj$ are calculated
with {\tt ALPGEN} \cite{alpgen}. The bottom quark mass $m_b=4.8$ GeV is kept
in all cases, and we take $M_H = 115$ GeV. We use structure functions CTEQ5L
\cite{cteq}, with $Q^2 = \hat s$  for $T \bar T$, $t \bar t$ and $t \bar b j$,
and $Q^2 = M_{W,Z}^2 + p_{T_{W,Z}}^2$ for $W b \bar b jj$, $Z b \bar b jj$, 
being $\sqrt{\hat s}$ the partonic centre of mass energy, and $p_{T_{W,Z}}$ the
transverse momentum of the gauge boson.\footnote{We find that $T \bar T$ cross
sections are $16-18$\% larger (for $m_T = 500-1000$ GeV) when evaluated with
MRST 2004 structure functions \cite{mrst} and their corresponding value of
$\alpha_s(M_Z)$. Assuming that background cross sections in the kinematical
region of
interest (large transverse momenta and invariant masses) scale by the same
rate, this would amount to a $8-9$\% increase in the statistical significance.}

The events are passed through {\tt PYTHIA 6.228} \cite{pythia} as external
processes to perform hadronisation and include initial and final state radiation
(ISR, FSR) and multiple interactions. We use the standard {\tt PYTHIA} settings
except for $b$ fragmentation, in which we use the Peterson parameterisation with
$\epsilon_b=-0.0035$ \cite{epsb}. A fast detector simulation {\tt ATLFAST 2.60} 
\cite{atlfast}, with standard settings, is used for the modelling of the ATLAS
detector. We reconstruct jets using a standard cone algorithm with
$\Delta R \equiv \sqrt{(\Delta \eta)^2 + (\Delta \phi)^2} = 0.4$, where $\eta$
is the pseudorapidity and $\phi$ the azimuthal angle. We do not apply trigger
inefficiencies and assume a perfect
charged lepton identification.
The package {\tt ATLFASTB} is used to recalibrate jet energies
and perform $b$ tagging, for which we select efficiencies of 60\%, 50\% for the
low and high luminosity LHC phases, respectively.

\section{Numerical results}
\label{sec:4}

The hadronised events are required to fulfill these two criteria: (a) the
presence of one (and only one) isolated charged lepton, which must have
transverse momentum $p_t \geq 20$ GeV and  $|\eta| \leq 2.5$; (b)
at least four jets with $p_t \geq 20$ GeV, $|\eta| \leq 2.5$, with exactly two
$b$ tags. The cross sections times efficiency of the five processes after these
pre-selection cuts are collected in Table~\ref{tab:1}, using a 60\% $b$
tagging rate. The events are produced without kinematical cuts at the
generator level in the case of $T \bar T$, $t \bar t$, while for the other
processes we set some loose cuts,
less restrictive than the ones used after hadronisation, which do not
bias the calculation. For $t \bar b j$
we only require pseudorapidities $|\eta| \leq 3$ for $b$, $j$. For
$W b \bar b jj$ we set
$p_t \geq 15$ GeV and $|\eta| \leq 3$ for the charged lepton, the $b$
quarks and the jets, and lego-plot separations $\Delta R_{jj},\Delta R_{bj},
\Delta R_{bb} \geq 0.4$, $\Delta R_{\ell b},\Delta R_{\ell j} \geq 0.2$. For
$Z b \bar b jj$ we require $p_t \geq 15$ GeV, $|\eta| \leq 3$ for $b$ quarks
and jets, $|\eta| \leq 10$ for the charged leptons, and
$\Delta R_{jj},\Delta R_{bj},\Delta R_{bb} \geq 0.4$.

\begin{table}[htb]
\begin{center}
\begin{tabular}{cc}
Process & $\sigma \times \mathrm{eff}$ \\
\hline
$T \bar T$ (500) & 44.9 fb \\
$T \bar T$ (1000) & 0.638 fb \\
$t \bar t$ & 18.8 pb \\
$W b \bar b jj$ & 1.23 pb \\
$Z b \bar b jj$ & 246 fb \\
$t \bar b j$ & 710 fb
\end{tabular}
\caption{Cross sections of the $T \bar T$ signal (with $m_T = 500,1000$ GeV)
and its backgrounds after pre-selection cuts.}
\label{tab:1}
\end{center}
\end{table}

\begin{figure}[p]
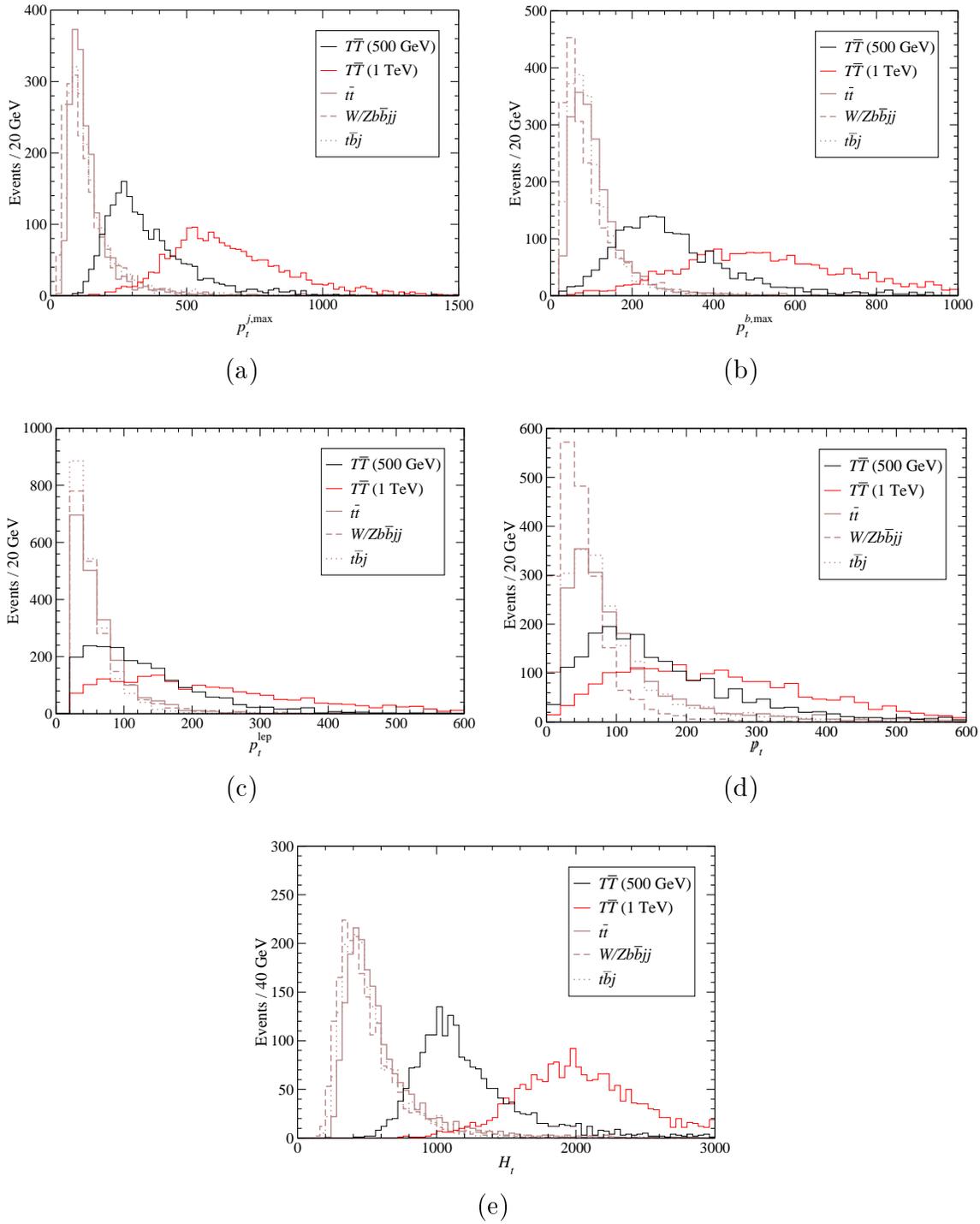

\begin{center}
\begin{tabular}{cc}
\epsfig{file=Figs/PTmax.eps,height=5.2cm,clip=} &
\epsfig{file=Figs/PTbmax.eps,height=5.2cm,clip=} \\
(a) & (b) \\[0.5cm]
\epsfig{file=Figs/PTlep.eps,height=5.2cm,clip=} &
\epsfig{file=Figs/PTmiss.eps,height=5.2cm,clip=} \\
(c) & (d) \\[0.5cm]
\multicolumn{2}{c}{\epsfig{file=Figs/HT.eps,height=5.2cm,clip=}} \\
\multicolumn{2}{c}{(e)}
\end{tabular}
\caption{Transverse momentum of: (a) the fastest jet; (b) the fastest $b$ jet;
(c) the charged lepton. Missing transverse momentum (d); total transverse
energy (e). The histograms are normalised to a
total number of 2000 events.}
\label{fig:1}
\end{center}
\end{figure}

The SM backgrounds are much larger than the signal, but
they concentrate in the low transverse momenta region. To reduce
them, it is useful to examine their dependence on the transverse momenta
of the charged lepton $\ptlep$, the fastest jet $\ptmax$ and the fastest $b$
jet $\ptbmax$, as well as the missing transverse momentum $\ptmiss$ and the
total transverse energy $H_t = \sum_{j,\ell,\gamma} p_t+\ptmiss$.
The kinematical distributions of these variables
are shown in Fig.~\ref{fig:1}. We display a weighted sum of the $W b \bar b jj$
and $Z b \bar b jj$ processes so as to reduce the number of histograms,
while the other backgrounds are shown separately.

\begin{figure}[htb]
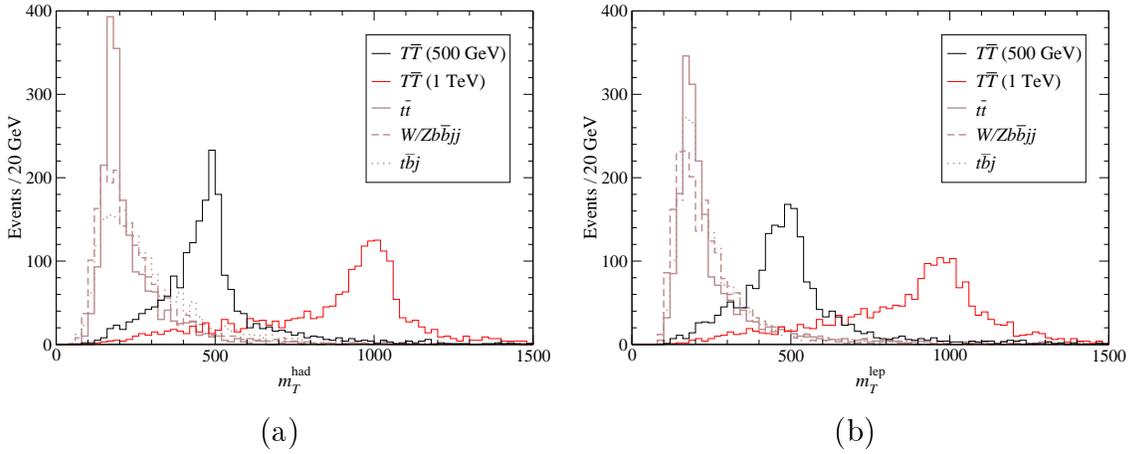

\begin{center}
\begin{tabular}{cc}
\epsfig{file=Figs/mT-had.eps,height=5.2cm,clip=} &
\epsfig{file=Figs/mT-lep.eps,height=5.2cm,clip=} \\
(a) & (b)
\end{tabular}
\caption{Reconstructed masses of the heavy quarks decaying hadronically (a) and
semileptonically (b). The histograms are normalised to a
total number of 2000 events.}
\label{fig:2}
\end{center}
\end{figure}

The $T \bar T$ signal can be discovered by the presence of peaks in the
invariant mass distributions corresponding to the two decaying quarks. In order
to reconstruct their momenta we first identify the two jets $j_1$,
$j_2$ from the $W$ decaying hadronically. The first one $j_1$ is chosen to be
the highest $p_t$ non-$b$ jet, and the second one $j_2$ as the non-$b$ jet
having with $j_1$ an invariant mass closest to $M_W$. The missing transverse
momentum is assigned to the undetected neutrino, and its longitudinal momentum
and energy are found requiring that the invariant mass of the charged lepton
and the neutrino is the $W$ mass, $(p_\ell+p_\nu)^2 = M_W^2$. This equation
yields two possible solutions. In addition, there are two different pairings of
the two $b$ jets to the $W$ bosons decaying hadronically and leptonically,
giving four possibilities for the reconstruction of the heavy quark momenta.
We select the one giving closest invariant masses $\mthad$, $\mtlep$ for the
quarks decaying hadronically and semileptonically. Their kinematical
distributions are shown in Fig.~\ref{fig:2}. In our calculations we have set
$V_{Tb} = 0.2,\, 0.1$ for $m_T = 500, \,1000$ GeV, respectively, yielding the
total widths $\Gamma_T = 2.80,\,6.16$ GeV . The cross
sections are independent of $V_{Tb}$ and for $\Gamma_T$ of these sizes the
broadness of the mass distributions too.

We point out that in our signal calculation we have not included other $T$
production processes giving the same experimental signature of one charged
lepton, four jets (with two $b$ tags) plus missing energy. Such processes do not
constitute a background (they are absent in the SM)
but instead increase the signal cross section.
One example is $T \bar T$ production in the decay channel
$T \bar T \to Z t \, W^- \bar b \to \nu \bar \nu W^+ b \, W^- \bar b$, with one
$W$ boson decaying hadronically and the other one leptonically. This process
has a cross section 10 times smaller than the one in Eq.~(\ref{ec:sig}). 
Other possible $T \bar T$ decay channels are
$T \bar T \to Z t \, W^- \bar b \to b \bar b W^+ b \, W^- \bar b$,
$T \bar T \to H t \, W^- \bar b \to b \bar b W^+ b \, W^- \bar b$ (assuming a
light Higgs boson) with two $b$ quarks mistagged. Their contributions
represent a $\sim 7$\% and $\sim 40$\% increase, respectively, in the total
cross section. Nevertheless, neither of the three processes mentioned yields
peaks in the $\mthad$, $\mtlep$ invariant mass
distributions, as reconstructed here for the $T \bar T \to W^+ b \, W^- \bar b$
signal, and their contributions are not likely
to be detectable due to the uncertainty in the SM background normalisation.
The same comments apply to $T \bar b j$ production with
radiation of an extra hard jet.

\subsection{Results for $\boldsymbol{m_T=500}$ GeV}

The large background cross sections make it convenient to introduce further
kinematical cuts at the generator level to reduce the number of events
processed with {\tt PYTHIA} and {\tt ATLFAST}. We require the presence of a
charged lepton with $p_t \geq 30$ GeV,
a jet with $p_t \geq 200$ GeV and, for $W b \bar b jj$ and $Z b \bar b jj$,
one $b$ quark with $p_t \geq 100$ GeV. This last cut does not bias the sample
because in these two processes the two non-$b$ jets mostly originate from light
quarks and gluons, for which the $b$ mistag probability is very low. Thus, the
$b$-tagged jets correspond to the $b$ quarks most of the time.
The kinematical cuts used to reduce backgrounds are
\begin{align}
& \ptmax \geq 250 ~\mathrm{GeV} \,, \quad
\ptbmax \geq 150 ~\mathrm{GeV} \,, \nonumber \\
& \ptlep \geq 50 ~\mathrm{GeV} \,, \quad
50 ~\mathrm{GeV} \leq \ptmiss \leq 600 ~\mathrm{GeV} \,, \nonumber \\
& H_t \geq 1000  ~\mathrm{GeV} \,.
\label{ec:cut500}
\end{align}
The cut $\ptmiss \leq 600$ GeV is useful because $t \bar t$ production with
large invariant masses is sometimes associated to very large $\ptmiss$, in
contrast with the signal. We also note that
with these requirements the charged lepton and the hardest jet provide a
trigger in the low luminosity LHC phase. The cross sections at the generator
level are listed in the first column of Table \ref{tab:500}, mainly for
informative purposes. The second column represents the number of events
$N_0 = K \sigma \mathcal{L}$ simulated, taking a luminosity of
10 fb$^{-1}$ and including the rescaling factors $K$ as explained in the
appendix. The figures in these two columns corresponding to different
processes should not be compared, since they are obtained with different initial
cuts in the event generation. 
Instead, the number of events $N_\mathrm{cut}$ surviving the selection criteria
in Eq.~(\ref{ec:cut500}) reflect the relative size of the processes after cuts.
They are shown in the third column. (The size of the signals and backgrounds
before cuts can be read from Table~\ref{tab:1}.)

\begin{table}[htb]
\begin{center}
\begin{tabular}{ccccc}
Process & $\tilde \sigma$ & $N_0$ & $N_\mathrm{cut}$ & $N_\mathrm{peak}$ \\
\hline
$T \bar T$ (500) & 204 fb & 2700 & 272 & 173 \\
$t \bar t$ & 5590 fb & 70000 & 1609 & 240 \\
$W b \bar b jj$ & 928 fb & 16000 & 287 & 65 \\
$Z b \bar b jj$ & 364 fb & 7200 & 39 & 10 \\
$t \bar b j$ & 626 fb & 8300 & 70 & 11 \\
\end{tabular}
\caption{For each process: cross sections $\tilde \sigma$ including cuts at the
generator level; number of events simulated $N_0$; number of events
$N_\mathrm{cut}$ passing the selection criteria in Eq.~(\ref{ec:cut500});
number of events $N_\mathrm{peak}$ passing the selection cuts which are in
the peak regions.}
\label{tab:500}
\end{center}
\end{table}

\begin{figure}[htb]
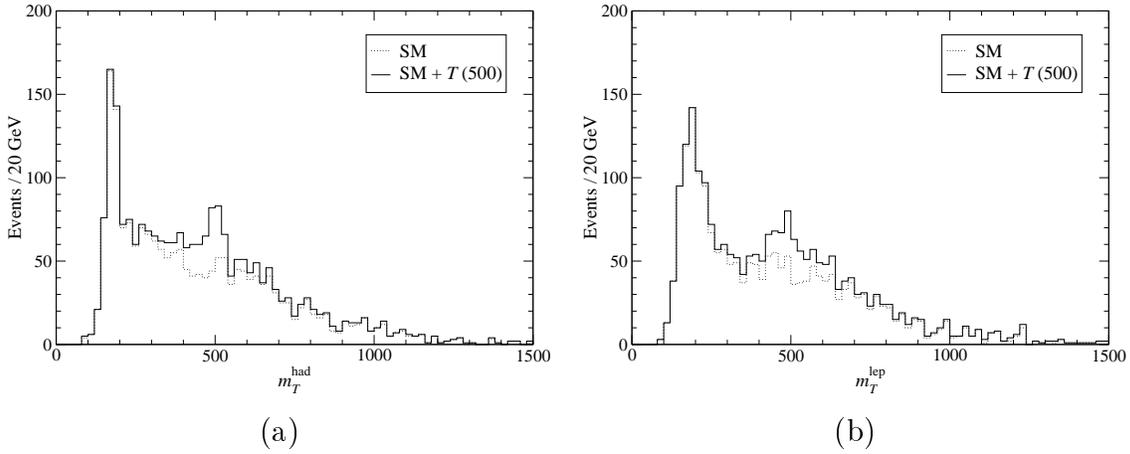

\begin{center}
\begin{tabular}{cc}
\epsfig{file=Figs/mT500-had.eps,height=5.2cm,clip=} &
\epsfig{file=Figs/mT500-lep.eps,height=5.2cm,clip=} \\
(a) & (b)
\end{tabular}
\caption{Reconstructed masses of the heavy quarks decaying hadronically (a) and
semileptonically (b), after the selection cuts in Eq.~(\ref{ec:cut500}). The
dashed lines correspond to the SM predictions, while the full lines represent
the SM plus a new 500 GeV quark.}
\label{fig:mT500}
\end{center}
\end{figure}

These kinematical cuts allow to detect the presence of the new quark in the
invariant mass distributions $\mthad$, $\mtlep$ as can be
observed in Fig.~\ref{fig:mT500}. The number of events in the peak regions
\begin{equation}
340 ~\mathrm{GeV} \leq \mthad \leq 660 ~\mathrm{GeV} \;, \quad
340 ~\mathrm{GeV} \leq \mtlep \leq 660 ~\mathrm{GeV}
\label{ec:peak500}
\end{equation}
are displayed in the fourth column of Table~\ref{tab:500}. They give a
statistical significance $S/\sqrt B = 9.6$. A $5\sigma$ significance, needed to
claim discovery, can be achieved with a luminosity $\mathcal{L} \simeq 2.7
~\mathrm{fb}^{-1}$. These numbers only consider statistical uncertainties,
assuming that the SM background can be normalised with the cross section
measurements ouside the peak region in Eq.~(\ref{ec:peak500}). Additionally, the
trigger and charged lepton detection efficiencies must be taken into account,
what reduces the statistical significance by a factor $\sim 0.95$.

\subsection{Results for $\boldsymbol{m_T=1}$ TeV}

\begin{figure}[htb]
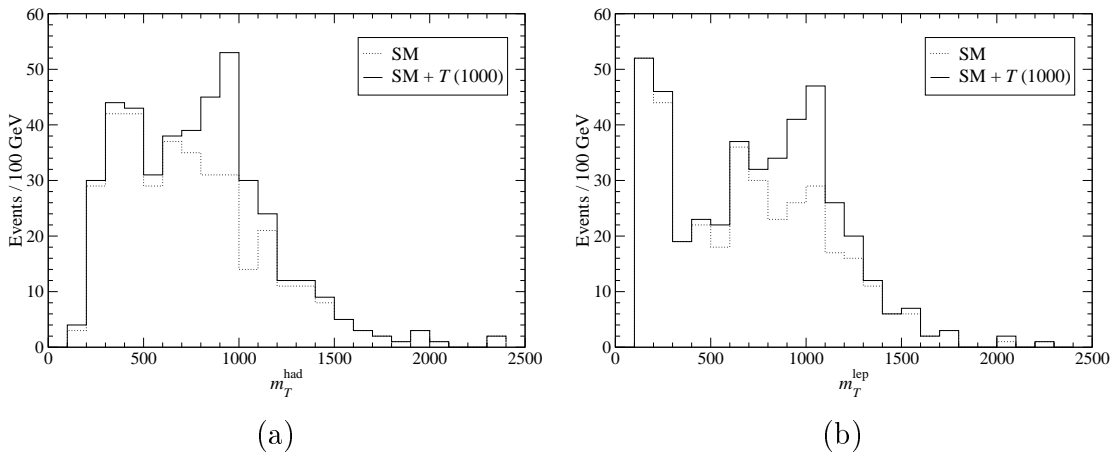

\begin{center}
\begin{tabular}{cc}
\epsfig{file=Figs/mT1000-had.eps,height=5.2cm,clip=} &
\epsfig{file=Figs/mT1000-lep.eps,height=5.2cm,clip=} \\
(a) & (b)
\end{tabular}
\caption{Reconstructed masses of the heavy quarks decaying hadronically (a) and
semileptonically (b), after the selection cuts in Eq.~(\ref{ec:cut1000}). The
dashed lines correspond to the SM predictions, while the full lines represent
the SM plus a new 1 TeV quark.}
\label{fig:mT1000}
\end{center}
\end{figure}

We repeat the same analysis for a heavy quark with $m_T=1$ TeV, in this case
choosing a $b$ tagging rate of 50\% at the high luminosity phase. The generator
cuts are raised to $p_t \geq 150$ GeV for the charged lepton, $p_t \geq 250$
GeV for the hardest jet and $p_t \geq 150$ GeV for the hardest $b$ quark, the
latter cut only for $W b \bar b jj$ and $Z b \bar b jj$ production. The
parton-level
cross sections for the five processes are listed in Table~\ref{tab:1000},
together with the number of simulated events $N_0$, corresponding to an
integrated luminosity of 300 fb$^{-1}$. The selection criteria
used to reduce backgrounds are
\begin{align}
& \ptmax \geq 400 ~\mathrm{GeV} \,, \quad
\ptbmax \geq 300 ~\mathrm{GeV} \,, \nonumber \\
& \ptlep \geq 200 ~\mathrm{GeV} \,, \quad
50 ~\mathrm{GeV} \leq \ptmiss \leq 400 ~\mathrm{GeV} \,, \nonumber \\
& H_t \geq 1800  ~\mathrm{GeV} \,.
\label{ec:cut1000}
\end{align}
With these cuts the charged lepton and hardest jet provide a trigger for the
event. The peak regions in this case are defined as
\begin{equation}
800 ~\mathrm{GeV} \leq \mthad \leq 1200 ~\mathrm{GeV} \;, \quad
800 ~\mathrm{GeV} \leq \mtlep \leq 1200 ~\mathrm{GeV} \,.
\end{equation}
The invariant mass distributions $\mthad$, $\mtlep$ after the cuts in
Eq.~(\ref{ec:cut1000}) are shown in Fig.~\ref{fig:mT1000}. The excess of events
in the peak regions amounts to $9.4$ standard deviations of the expected SM
background. A $5\sigma$ significance would be achieved with an integrated
luminosity $\mathcal{L} \simeq 85~\mathrm{fb}^{-1}$. With a simple rescaling it
can be estimated that masses up to $m_T = 1.1$ TeV can be discovered with
$5\sigma$ significance for $\mathcal{L} = 300 ~\mathrm{fb}^{-1}$ and, if no
signal is found, the 95\% CL limit $m_T \geq 1.3$ TeV can be set.

\begin{table}[htb]
\begin{center}
\begin{tabular}{ccccc}
Process & $\tilde \sigma$ & $N_0$ & $N_\mathrm{cut}$ & $N_\mathrm{peak}$ \\
\hline
$T \bar T$ (1000) & 2.89 fb & 1330 & 70 & 48 \\
$t \bar t$ & 778 fb & 294000 & 208 & 10 \\
$W b \bar b jj$ & 66.8 fb & 34000 & 132 & 15 \\
$Z b \bar b jj$ & 48.0 fb & 28500 & 19 & 1 \\
$t \bar b j$ & 44.1 fb & 17500 & 3 & 0 \\
\end{tabular}
\caption{For each process: cross sections $\tilde \sigma$ including cuts at the
generator level; number of events simulated $N_0$; number of events
$N_\mathrm{cut}$ passing the selection criteria in Eq.~(\ref{ec:cut1000});
number of events $N_\mathrm{peak}$ passing the selection cuts which are in the
peak regions.}
\label{tab:1000}
\end{center}
\end{table}

\section{Summary and discussion}
\label{sec:5}

Up-type quark singlets with charge $Q=2/3$ are predicted in some SM
extensions, with masses ranging from few hundreds of GeV to several TeV.
Their observation would represent not only a clear new physics signal but also
an important confirmation for these models. We have shown that for $m_T =
500$ GeV a $5\sigma$ statistical significance would be attained already with 
3 fb$^{-1}$ of integrated luminosity, which can be collected after few months of
LHC operation in its first phase. For new quarks in this mass range, $T \bar T$
production provides the best signal of their presence, allowing a prompt
discovery if they exist. With an integrated luminosity of 300 fb$^{-1}$,
$T \bar T$ production may discover a new quark with $m_T \leq 1.1$ TeV, or set
a 95\% CL bound $m_T > 1.3$ TeV, independent of $V_{Tb}$, if no signal is
observed.

We point out that if a fourth quark generation $(T,B)$ exists with
$m_T < m_B$ the dominant decay of the up-type quark is $T \to W b$, giving the
same signal studied here. ($T \to Zt ,\, Ht$ are forbiden at the tree
level by the vanishing of the flavour-changing neutral coupling $X_{tT}$.)
The results obtained for an up-type singlet can then be straightforwardly
applied to a fourth generation quark, multiplying the statistical significances
in section \ref{sec:4} by a factor of four. If $m_T < m_B$, a fourth
generation quark $T$ with $m_T \leq 1.3$ TeV could be discovered with 300
fb$^{-1}$ of integrated luminosity, and the 95\% CL bound $m_T > 1.5$ TeV could
be set if they are not observed. If $B$ is lighter than $T$,
the decay $T \to W^+ B$, with $W$ on its mass shell if $m_T > m_B+M_W$, is open.
Hence, the branching ratio of $T \to W b$ depends on $V_{Tb}$, and
model-independent predictions cannot be made. We also remark that
in case that a new quark $T$ is discovered without a $Q=-1/3$
partner, the experimental search for the decays $T \to Zt ,\, Ht$ can
determine if $T$ is a $\mathrm{SU}(2)_L$ singlet or belongs to a doublet (in
which case its partner $B$ should be heavier and undetected).

\begin{figure}[htb]
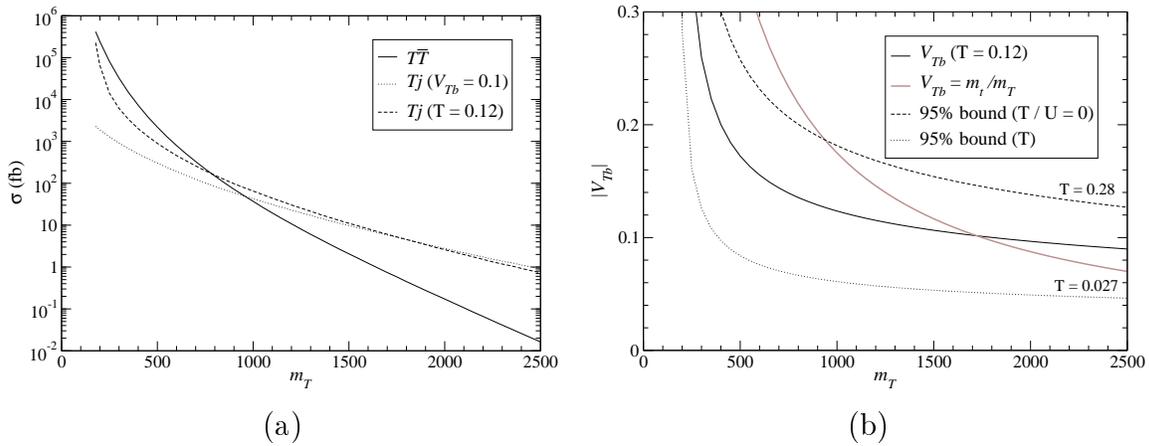

\begin{center}
\begin{tabular}{cc}
\epsfig{file=Figs/mass-cross.eps,height=5.12cm,clip=} &
\epsfig{file=Figs/Vguess.eps,height=5.12cm,clip=} \\
(a) & (b)
\end{tabular}
\caption{(a) Cross sections for $T \bar T$ production (full line) and $Tj$
production, in the latter case for $V_{Tb}=0.1$ (dotted line), and for $V_{Tb}$
derived from the $\mathrm{T}$ parameter (dashed line). (b) Upper bounds on
$|V_{Tb}|$ and values suggested by the $\mathrm{T}$ parameter (black line) and
the Little Higgs relation $V_{Tb} = m_t/m_T$ (grey line).}
\label{fig:cs}
\end{center}
\end{figure}

New $Q=2/3$ quark singlets can be produced in association with light jets as
well, mainly in the processes $u b \to d T$, $d \bar b \to u \bar T$. The cross
sections for $Tj$ production are quadratic in $|V_{tb}|$, but on the other hand
they do not decrease with $m_T$ as quickly as
for $T \bar T$. We plot in Fig.~\ref{fig:cs} (a) the cross sections for
$T \bar T$ and $Tj$ production for different heavy quark masses. 
For $Tj$ we select a fixed coupling $V_{Tb} = 0.1$ as well as an $m_T$-dependent
coupling suggested by the experimental central value $\mathrm{T} = 0.12$
(obtained for $\mathrm{U}=0$) and Eq.~(\ref{ec:T}). The $V_{Tb}$ values derived
from
$\mathrm{T} = 0.12$ are shown in Fig.~\ref{fig:cs} (b). Of course, in models
beyond the SM there may be additional contributions to oblique corrections,
thus we take the experimental measurement of $\mathrm{T}$ only as a hint on
the size of $V_{Tb}$. In Fig.~\ref{fig:cs} (b) we also plot the Little Higgs
model relation $V_{Tb} = m_t/m_T$ and the two 95\% CL upper limits on
$|V_{Tb}|$ obtained from
\begin{align}
& \mathrm{T} \leq 0.28 & & (95\% ~\mathrm{CL}\,, ~ \mathrm{U}=0) \,,
  \nonumber \\
& \mathrm{T} \leq 0.027 & & (95\% ~\mathrm{CL}\,, ~ \mathrm{U~arbitrary}) \,,
\label{ec:Tlim}
\end{align}
respectively.
We note that that the assumption $V_{Tb} = m_t/m_T$ potentially conflicts
with the $\mathrm{T}$ parameter measurement for $m_T < 900$ GeV, even using the
less restrictive bound $T \leq 0.28$.

\begin{figure}[htb]
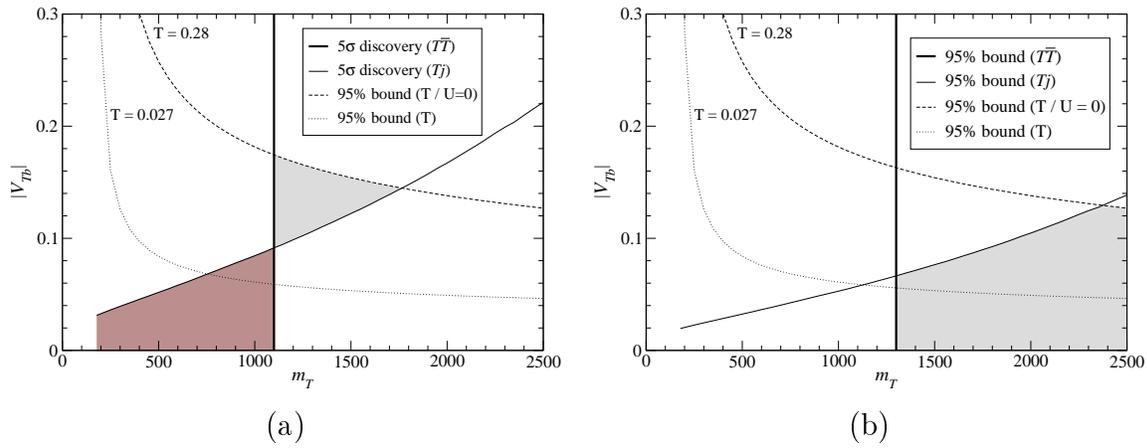

\begin{center}
\begin{tabular}{cc}
\epsfig{file=Figs/V5s.eps,height=5.09cm,clip=} &
\epsfig{file=Figs/V2s.eps,height=5.09cm,clip=} \\
(a) & (b)
\end{tabular}
\caption{(a) $5 \sigma$ discovery limits for the new quark (full lines), and
indirect bounds from the $\mathrm{T}$ parameter, explained in the text.
(b) Combined 95\% CL bounds on $m_T$, $|V_{Tb}|$ (shaded area) from
$T \bar T$, $Tj$ production and the $\mathrm{T}$ parameter, if $Q=2/3$ quark
singlets are not observed at LHC.}
\label{fig:limits}
\end{center}
\end{figure}

The discovery potential of $Tj$ production can be estimated from existing
analyses. This process, with $T \to Wb \to \ell \nu b$, gives a $21.5\sigma$
significance for $m_T = 1$ TeV and $V_{Tb} = m_t/m_T \simeq 0.175$
\cite{costanzo}. We make the reasonable assumption that for different $T$
masses the signal to background ratio $S/B$ in the kinematical region of
interest (with high $p_t$ and invariant masses $\sim m_T$) remains approximately
constant (obviously keeping equal CKM factors $|V_{Tb}|^2$ for the
signal).\footnote{This assumption is justified by the decrease of the
tails in the transverse momenta and invariant mass distributions of the SM
background. More optimistic extrapolations of the SM background cross
section, {\em e.g.} assuming that it decreases faster than the $Tj$ signal,
would lead to higher discovery limits on $T$ masses, as the ones given in
Ref.\cite{azuelos}.}
Requiring a statistical significance $S/\sqrt B = 5$
sets a lower limit on the coupling $V_{Tb}$ for each $m_T$ value. These limits
are plotted in Fig.~\ref{fig:limits} (a), together with the discovery
limit $m_T \leq 1.1$ TeV for $T$ pair production. We also include the 95\% CL
bounds from the $\mathrm{T}$ parameter in Eq.~(\ref{ec:Tlim}). We point out that
if $\mathrm{T} \leq 0.027$ is enforced the discovery reach of $T \bar T$
production is higher than for $Tj$, since for $m_T \gtrsim 700$ GeV
the $V_{Tb}$ values needed for $5\sigma$ discovery in $Tj$ production are
not allowed. Assuming the less restrictive limit $\mathrm{T} \leq 0.28$, there
is a region (light shaded area in the figure) where the new
quark can be discovered in single $T$ but not in $T$ pair production.
Conversely, in the dark shaded area the new quark can be discovered in $T \bar
T$ production but not in single $T$ processes.

The discovery of a new $Q=2/3$ quark singlet would certainly be a rather
important achievement towards the understanding of the flavour structure of the
SM, and might help explain the largeness of the top quark mass
\cite{paco,lhiggs}. On the other hand,
the non-observation of new quarks at LHC would also be interesting
on its own. In such case, the combined bounds obtained from single and $T$ pair
production and the $\mathrm{T}$ parameter would restrict $m_T$, $|V_{Tb}|$ to
lie in the shaded area in Fig.~\ref{fig:limits} (b). (If we use the more
restrictive bound in Eq.~(\ref{ec:Tlim})
the allowed region is somewhat smaller, as can be seen in the figure.) In this
area we have $m_T \geq 1.3$ TeV and $|V_{Tb}| \leq 0.13$, the latter implying
$|V_{tb}| \geq 0.991$. For $m_T \geq
600$ GeV, the couplings $V_{Td}$, $V_{Ts}$ are already very constrained by kaon
and $B$ physics measurements \cite{largo}. Therefore, the non-observation of a
new quark would significantly improve the indirect limits on CKM matrix
elements $V_{td}$, $V_{ts}$, $V_{tb}$ within this class of models.

\vspace{1cm}
\noindent
{\Large \bf Acknowledgements}

\vspace{0.4cm} \noindent
I thank F. del Aguila and R. Pittau for useful discussions and a critical
reading of the manuscript. This work has been supported by FCT through project
CFTP-FCT UNIT 777 and grant SFRH/BPD/12603/2003.

\appendix
\section{Appendix: signal and background normalisation}
\label{sec:a}

In addition to the processes listed in Eqs.~(\ref{ec:sig}),(\ref{ec:bkg}) there
are other higher order processes contributing to the signal and backgrounds,
namely final states including extra jets from QCD radiation. Indeed,
$t \bar b j$ production only
yields three jets at the partonic level, and the fourth one required to pass our
pre-selection criteria must be originated by radiation.
These higher-order processes are approximately accounted for by {\tt PYTHIA}
showering, which generates hard extra jets by FSR. For example,
in $W b \bar b jj$ production some fraction of the
events, when passed through {\tt PYTHIA} showering, are converted into
$W b \bar b jjj$ or $W b \bar b jjjj$
events, the additional jets with a high ($\gtrsim 50$ GeV) transverse momentum.

One possible method to take higher order processes into account is to
generate them at the parton level, forbidding {\tt PYTHIA} to radiate hard extra
jets to avoid double counting \cite{presc} but allowing the soft and collinear
ones. Instead, our approach is to normalise the numbers of events simulated
$N_0$
by approximate correction factors $K$ so that these figures correspond to the
processes in  Eqs.~(\ref{ec:sig}),(\ref{ec:bkg}) plus higher order ones.
The number of events simulated for a given process is then
$N_0 = K \sigma \mathcal{L}$, being $\sigma$ the cross section of the process
and $\mathcal{L}$ the luminosity. The correction factor $K$ is calculated in
the $W b \bar b jj$ example as follows:
\begin{enumerate}
\item We generate $W b \bar b jj$ events requiring high
transverse momenta $p_t \geq 100$ GeV for jets at the generator level,
and a large separation $\Delta R \geq 0.6$ among all partons.
\item These events are passed twice through {\tt PYTHIA} and {\tt ATLFAST},
including FSR and without including it. In both cases ISR, multiple interactions
and energy smearing are turned off, because we want to isolate the effect of
FSR.
\item We examine the number of events with four-jets (corresponding to the four
initial partons) with $p_t \geq 80$ GeV in both samples. Let us call these
numbers $n_4^{0}$, $n_4^{\mathrm{FSR}}$, respectively. $K$ is then defined as
the ratio between them, $K = n_4^{0} / n_4^{\mathrm{FSR}}$.
\end{enumerate}
With this rescaling factor definition, the number of four-jet events after FSR
corresponds to the one present at the parton level.
The value of $K$ depends on the cut values for the event generation (100 GeV)
and the jet counting (80 GeV), and therefore this procedure is approximate.
However, for our purposes this simple and fast $K$-factor prescription seems
sufficient. For the
$T \bar T$ signal we obtain $K(T \bar T) \simeq 1.3$, $K(T \bar T) \simeq 1.5$
for heavy masses of 500 GeV and 1 TeV, respectively, and for the backgrounds
$K(t \bar t) \simeq 1.3$, $K(t \bar b j) \simeq 1.3$,
$K(W b \bar b jj) \simeq 1.7$, $K(Z b \bar b jj) \simeq 2.0$.

\end{document}